\documentclass{aa}
\usepackage{graphicx}

\newcommand{\RM}{\mbox{RM}}

\begin{document}

   \title{Canals Beyond Mars: Beam Depolarization in Radio Continuum
   Maps of the warm ISM} 

   \titlerunning{Beam Depolarization in the warm ISM}

   \author{M. Haverkorn\inst{1,2}
          \and
          F. Heitsch\inst{3,4,5}}

   \offprints{M. Haverkorn}

   \institute{Sterrewacht Leiden
              P.O. Box 9513, NL-2300 RA Leiden, The Netherlands 
              \and
              Harvard-Smithsonian Center for Astrophysics, 
              60 Garden Street MS-67, Cambridge, MA, 02138, USA\\
              \email{mhaverkorn@cfa.harvard.edu}
         \and
             Max-Planck-Institut f\"ur Astronomie,
             K\"onigstuhl 17, 69117 Heidelberg, Germany
         \and
             University of Wisconsin-Madison,
             475 N Charter St, Madison, WI 53706, USA
         \and 
             Institut f\"ur Astronomie und Astrophysik, Universit\"at
              M\"unchen, Scheinerstr. 1, 81679 M\"unchen, Germany\\
             \email{heitsch@usm.uni-muenchen.de}
             }
   \date{}

   \abstract{Multi-frequency radio polarimetric observations of the
             diffuse Galactic synchrotron background enable us to 
             study the structure of the diffuse ionized gas via 
             rotation measure maps. However, depolarization will
             introduce artifacts in the resulting rotation measure,
             most notably in the form of narrow, elongated
             ``depolarization canals''. We use numerical models
             of a non-emitting Faraday rotating medium to study the RM
             distribution needed to create depolarization canals by
             depolarization due to a finite beam width, and to
             estimate the influence of this depolarization mechanism
             on the determination of RM. We argue that the
             depolarization canals indeed can be caused by beam
             depolarization, which in turn is a natural consequence
             when observing a turbulent medium with limited
             resolution. Furthermore, we estimate that beam
             depolarization can induce an additional error of about
             20\% in RM determinations, and considerably less in
             regions that are not affected by depolarization canals. 
   }

   \maketitle
%
%________________________________________________________________

\section{Introduction}

The Galactic magnetic field and turbulence play a major r\^{o}le in 
structuring the Galactic disk. The recognition of the dynamical
importance of turbulence in the interstellar medium (ISM)
(e.g. Armstrong et al. \cite{ARS1995}) let our picture of the ISM
evolve from a simple three-phase gas to a highly complex
medium. Turbulence is believed to play a crucial r\^{o}le for
molecular cloud and star formation (see e.g. Larson \cite{LAR1981};
Mac Low \& Klessen \cite{MAK2003}), however, its origin is widely
debated. On a larger scale turbulence leads  to chemical mixing (de
Avillez \& Mac Low \cite{DAM2002}) and contributes to the vertical
pressure balance in the Galactic disk and to the heating of the ISM
(Minter \& Balser \cite{MAB1998}).

Magnetic fields are tightly linked to the turbulent nature of the ISM.
Their large-scale components help confining cosmic rays to the 
Galaxy, and thus affect chemical processes in molecular clouds. 
They probably contribute to generating turbulence in the disk 
(Sellwood \& Balbus \cite{SEB1999}) and definitely affect the vertical
disk structure. The ratio between mass and magnetic flux in the ISM
is an essential piece of information necessary to understand the 
r\^{o}le of magnetic fields for molecular cloud and eventually star 
formation (see e.g. Crutcher \cite{CRU1999}). 

Density and magnetic field structure in the ISM can be probed using
multi-frequency radio polarimetry to study Faraday 
rotation\footnote{Faraday rotation is the rotation of the plane of 
  linear polarization of radiation when it propagates through a
  magneto-ionic medium, due to the birefringence of the medium to
  circularly polarized radiation. The rotation of the polarization
  plane through the medium is $\Delta\phi = RM \lambda^2$, with
  rotation measure $RM = 0.81\int B_{\parallel}n_eds$, where
  $B_{\parallel}$ is the component of the magnetic field parallel to
  the line of sight in $\mu$G, $n_e$ is the thermal electron density
  in cm$^{-3}$ and $ds$ is the path length in pc.
}  
in the magneto-ionic medium. Polarized extragalactic  point sources and
pulsars have been used to determine rotation measures (RM), but
these give only measurements at certain lines of sight. Observing the
diffuse Galactic synchrotron background instead yields structure in
polarized emission on many scales, along many contiguous lines of
sight. However, depolarization complicates the interpretation of
the resulting RM maps. Depolarization occurs if the telescope
beam is larger than the scale of structure in polarization (beam
depolarization), along the line of sight if the radiation is emitted
and Faraday rotated largely in the same medium (depth depolarization
or internal Faraday dispersion), or if the frequency bandwidth is so
large that the polarization angle changes significantly within the
band (bandwidth depolarization), see e.g. Burn (\cite{BUR1966}) or
Sokoloff et al.\ (\cite{SBS1998}). In polarization observations of the
Galactic synchrotron background, often it is difficult if not
impossible to estimate the separate contributions of beam and depth
depolarization.

Depolarized regions often exhibit canal-like features
(e.g Duncan et al.\ \cite{DHJ1997}, \cite{DRR1999}, Gray
et al.\ \cite{GLD1999}, Uyan\i ker et al.\ \cite{UFR1999}, 
Haverkorn et al.\ \cite{HKB2000}, Gaensler et al.\ \cite{GDM2001}).
Both beam and depth depolarization can cause the canals, but
require a different underlying RM distribution. The goal of this paper
is to determine if the observed depolarization canals can be due to
beam depolarization, and how beam depolarization affects the computed
RM. To this effect, we use numerical models of the warm magneto-ionic
ISM to compare to observations of the diffuse Galactic synchrotron
background. The models are irradiated with uniform synchrotron
background emission, which is Faraday rotated in the modeled
medium. Then, the polarized radiation is smoothed to simulate a finite
telescope beam, where after RM's are computed from the smoothed
emission maps. The structure in polarized intensity and RM emerging
from the smoothed modeled radiation field is then compared to the
observations. This allows us to simulate purely the effect that beam
depolarization has on polarimetric observations of the ionized ISM.

This work was motivated by the observations briefly presented in
Sect.~\ref{s:obs}. In Sect.~\ref{s:nummod} we describe the numerical
models, the addition of the synchrotron background and the simulation
of beam depolarization. Section~\ref{s:beamdepol} discusses the
signature of beam depolarization in the maps of polarized intensity,
while in  Sect.~\ref{s:error} the effect of beam depolarization on RM
is studied.

%***************************************
\begin{figure}
  \centering
  \includegraphics[width=0.5\textwidth]{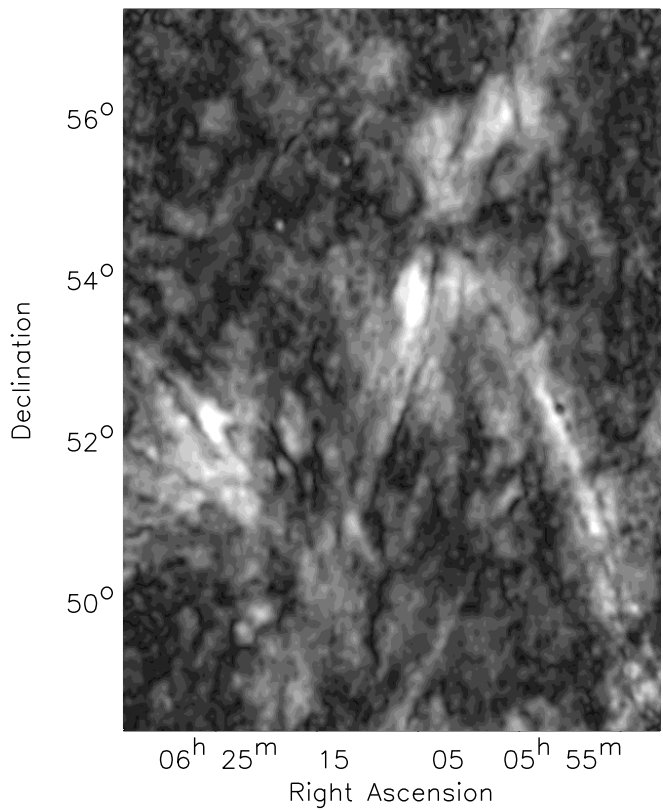}
  \caption{Observed polarized intensity at 349~MHz in a field centered
  at $(l,b) = (161\degr, 16\degr)$. High intensity is shown white, up to
  a maximum of $\sim$~13~K.} 
  \label{f:obs_p}
\end{figure}
%****************************************
%***************************************
\begin{figure}
  \centering
  \includegraphics[width=0.5\textwidth]{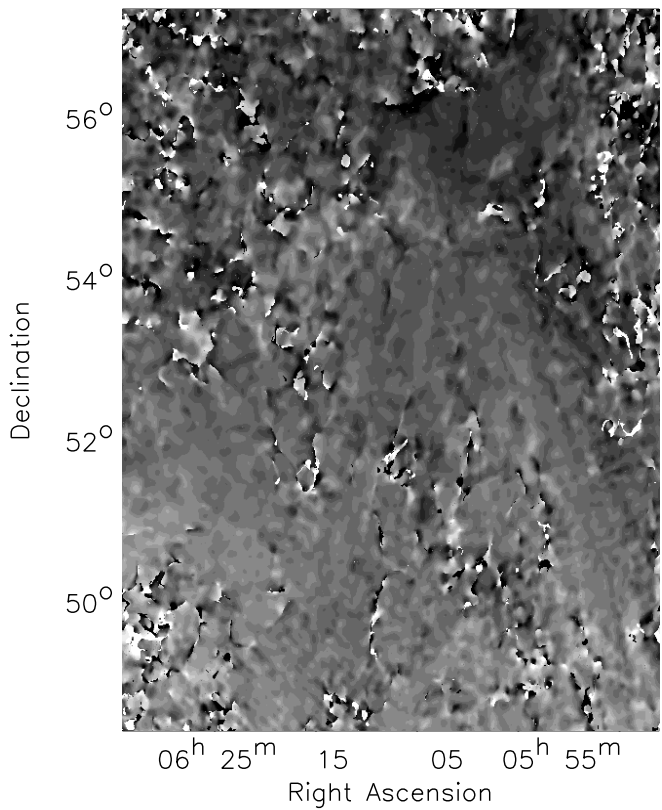}
  \caption{Map of RM in the observed field. The grey scale runs from
  $-15$ to 15~rad~m$^{-2}$. The map shows an RM value at every
  position, regardless of the quality of the linear fit of $\phi$
  against $\lambda^2$.}
  \label{f:obs_rm}
\end{figure}
%****************************************

%====================================================
\section{The observations}
\label{s:obs}
The observations were made with the Westerbork Synthesis Radio
Telescope (WSRT) in five frequency bands centered on 341, 349, 355,
365, and 375~MHz, each with a bandwidth of 5~MHz. The field discussed
here is centered at $(l, b) = (161\degr, 16\degr)$ and about 50 square
degrees in size, and has a resolution of about 4$\arcmin$. 
All four Stokes parameters $I$, $Q$, $U$ and $V$ are measured, from
which polarized intensity $P$ and polarization angle $\phi$ were
determined as
\begin{eqnarray}
  P    &=& \sqrt{Q^2 + U^2},\label{e:polint}\\
  \phi &=& \frac{1}{2}\arctan\frac{U}{Q}. \label{e:polang} 
\end{eqnarray}
The instrumental linear polarization is $\la 1\%$. 
For a detailed description of this field see Haverkorn et al.\
(\cite{HKB03a}). 

No signal was detected in total intensity $I$. As the interferometer
is only sensitive to scales smaller than approximately a degree, the
non-detection of $I$  means that $I$ does not exhibit any structure on
scales smaller than $\sim 1\degr$, so that the observed total
synchrotron power must be very uniform. On the other hand, abundant
structure in linearly polarized emission was detected, both in
$P$ (see Fig.~\ref{f:obs_p}) and in $\phi$. This indicates that the
small-scale structure in polarization is caused by Faraday rotation
and depolarization. Ubiquitously present in the field are one-beam
wide ``canals'' of beam depolarization, which we will discuss in
Sect.~\ref{s:beamdepol}. Missing large-scale structure in
$P$ probably does not play  a large r\^ole in this
field (Haverkorn et al.\ 2003a, 2004b).

RMs were computed from linear fits of $\phi$ against $\lambda^2$. For
about 70\% of the beams with an average polarized intensity $\langle
P\rangle > 5$~S/N, reliable RM's (with reduced $\chi^2 < 2$) could be
determined. This is about 28\% of all data. Fig.~\ref{f:obs_rm}
shows a grey scale map of RM, where in each beam a RM value is
plotted, regardless of the quality of fit of the
$\phi(\lambda^2)$-relation. In general, the positions showing
anomalous RM with rapidly changing magnitude and sign are the
positions where noise dominates, so that reliable RM determination
is not possible, see Sect.~\ref{s:beamdepol}.

Contributions to the RM are made along the line of sight at positions
with a non-zero electron density and magnetic field. This is the
case in the warm and hot ISM  components. The filling factor in the
warm ISM is approximately 20\% (Reynolds \cite{REY1991}), while the
electron density in the hot ISM is so low that its RM contribution can
be neglected for our purposes. Furthermore, polarized emission  is
increasingly depolarized when emitted at greater distances. Haverkorn
et al.\  (\cite{HKB04a}) estimate a distance of $\sim$~600~pc between
observer and the "polarization horizon", defined as the point beyond
which less than about a third of the polarized emission is not
depolarized while propagating through the medium.

%====================================================
\section{Numerical models\label{s:nummod}}

In order to estimate the effect of beam depolarization on the 
RM determinations, we generate artificial maps of polarized 
intensity at five wavelengths from numerical models of a turbulent
magnetized ISM.

\subsection{Domain considerations}

Ideally, the numerical models should resemble the observational
domain (i.e., the ratio between the length of the line of sight and
the extent in the plane of sky) while retaining a resolution in the
plane of sky comparable to the observations. At an estimated
polarization horizon of 600~pc, the spatial extent of
the $7\degr\times 9\degr$ field of observation is about 75~pc. 
The observational resolution of 4$\arcmin$ would correspond
to approximately $120\times 120$ resolution elements in the plane
of sky.

However, for the numerical models, this estimated requirement
is too optimistic for two reasons. First, in order to investigate
beam depolarization, we need substructure within one beam
(i.e. resolution element). Second, any magnetohydrodynamics (MHD) code
suffers from numerical dissipation at the smallest scales, thus
introducing a lower limit for physically reasonable structures in the
simulations (between 2 to 4 resolution elements in our case). Both
requirements taken together lead to a desirable resolution of $\sim
500\times 500$ elements in  the plane of sky for the simulations,
resulting in a domain size of $500\times 500\times 4000$
elements. This is beyond current computer capacities. 

Thus we decided to use a cubic simulation domain (at $512^3$ elements)
and to construct a bar-like domain by stacking this cube for 8 times, 
each time rotating it and shifting it by one fifth of the domain size. 
Thus we get different realizations of the turbulent  structures,
mimicking a turbulent medium over the whole length of the  line of
sight.

\subsection{Model description}

Our ``observational'' domain is derived from a periodic-box-simulation
of the turbulent magnetized ISM  by Li et al.\
(\cite{LNM2000,LNM2004}) and Heitsch et al.\ (\cite{HZM2001}),
employing ZEUS-MP, the massively parallel version of ZEUS-3D (Norman
\cite{NOR2000}). The electron density and the magnetic field are
rescaled to the values for the warm ISM, from which five models
with different path lengths through the medium were constructed, as
shown in Table~\ref{tab:nummodels}.

The density and the magnetic field are initially uniform, and
will be perturbed by a fixed velocity field. We generate the velocity
field as described by Mac Low (\cite{MAC1999}), as follows. Each
wave number $k$ is given an amplitude in Fourier space, which is
randomly drawn  from a Gaussian distribution around unity, and a
random phase. Fourier transforming to velocity space yields velocity
components in three dimensions for each position. This is the velocity
field with which the model is perturbed at each timestep, to simulate
turbulent driving. The velocity  components are multiplied by a
velocity amplitude at each timestep, chosen such that the input energy
stays constant with time. Perturbations are only induced at the lowest
spatial wavenumbers $1<k<2$. In the plane of sky, this corresponds to
the largest scales possible in the domain. The driving mechanism
is meant to mimic energy input on the largest scales by an
unspecified physical process such as supernova shock fronts or
Galactic shear.

We use a snapshot of the simulation at a time when the full turbulent
cascade has developed and a steady state between energy input on the
largest scales and numerical dissipation on the smallest scales has
been reached. This steady state corresponds to a Mach number 
of ${\cal M} \approx 10$. This is more than the values usually
assumed for the warm ISM (namely ${\cal M} \approx 1$ at a sound speed
of $\approx 6\mbox{ km}\mbox{ s}^{-1}$), thus rendering the
turbulence more compressive than intended for this work.

However, the turbulent cascade represented by a Kolmogorov slope
of $-5/3$ does not change substantially (see e.g. Cho et al.\
\cite{CLV2003}),  so that we can rescale the densities and field
strengths to values appropriate for the warm ISM. Fluctuations in
electron density are allowed up to an upper limit of $10$\% of the
mean density (assumed to be $0.1$~cm$^{-3}$), which corresponds to the
upper limit to structure in emission measure (EM) in the field of
observation given by high-resolution H$\alpha$ observations (Madsen,
private communication). The field variations are on the order of the
mean field of $2.5\,\mu$G.

We assume a filling factor of ionized gas of 100\%, and a sharp cutoff
of Faraday rotation of the polarized emission at 600~pc. As the RM is
an integral over the line of sight, a smaller filling factor combined
with a longer path length would give the same results.

The MHD model does not contain synchrotron emission within the
medium itself. This is obviously not a realistic situation, but allows
us to investigate the influence of beam depolarization without depth
depolarization, as the latter only occurs in a medium that both
Faraday rotates and emits. Including synchrotron emission in the
medium would result in additional depolarization and larger deviations
from the linear $\phi(\lambda^2)$ relation.

\begin{table}[h]
  \begin{tabular}{cccccc}
    \hline
    \hspace*{-0.3cm} name  & $N_x \times N_y \times N_z$ 
          & $L_{los}$
          & $B_{\|}$ 
          & $n_e$ 
          & $\Delta$RM \\
          & 
          & [pc] 
          & [$\mu$G]
          & \hspace*{-0.2cm}[cm$^{-3}$]
          & \hspace*{-0.2cm}[rad m$^{-2}$] \\
    \hline
    $B1$ & $512\times 512\times 4096$ & $ 100$&$2.5$&$0.1$& $ 3$  \\
    $B2$ & $512\times 512\times 4096$ & $ 200$&$2.5$&$0.1$& $ 6$  \\
    $B3$ & $512\times 512\times 4096$ & $ 400$&$2.5$&$0.1$& $10$  \\
    $B4$ & $512\times 512\times 4096$ & $ 600$&$2.5$&$0.1$& $15$  \\
    $B5$ & $512\times 512\times 4096$ & $ 800$&$2.5$&$0.1$& $25$  \\
    \hline
  \end{tabular}
  \caption{Parameters of the numerical models used: $N_x$, $N_y$
  and $N_z$ are the model dimensions in resolution elements and
  $L_{los}$ is the length of the line of sight in parsecs. The initial
  magnetic field $B_{\parallel}$ is oriented along the line of sight
  $z$ and $n_e$ is the initial uniform electron density. $\Delta$RM is
  the width $\sigma$ of the Gaussian distribution of RM, scaled via
  $L$.}
  \label{tab:nummodels}
\end{table}

\subsection{Simulating beam depolarization\label{ss:beamsim}}

The goal is to produce maps of Stokes $Q$ and $U$ from the models,
which then can be treated like observational data. From the bars we
construct the numerical RM in the plane of sky as 
\begin{equation}
  \RM_{i,j} = 0.81\,\frac{L_{los}}{N_z}\,\sum_{k=1}^{N_z} 
                    [n_e\,B_{\parallel}]_{i,j,k},\,\,\,\,
                    i,j\in\{1\cdots N_x\},
\end{equation}
in units of rad~m$^{-2}$. The line-of-sight length $L_{los}$ is given
in parsecs, the electron density $n_e$ in cm$^{-3}$, and the magnetic
field component along the line-of-sight $B_{\parallel}$ in
$\mu$G. $N_x, N_z$ are the number of cells in $x$ and $z$ direction,
respectively, and $N_y = N_x$. For simplicity, we integrate
along parallel lines of sight instead of an integration over a
cone-like domain. Stokes $Q_0$ and $U_0$ are derived from those maps
via 
\begin{eqnarray}
  \phi &=& \RM\,\lambda^2,\\
  U_0  &=& \sin(2\phi),\label{e:origstokesu}\\
  Q_0  &=& \cos(2\phi).\label{e:origstokesq}
\end{eqnarray}
Note that we assume $100$\% uniform background polarization, $P_0 =
\sqrt{Q_0^2 + U_0^2} = 1$, and a constant background polarization angle
of $\phi_0 = 0$. The wavelength $\lambda$ is given in meters.

To mimic the limited telescope resolution, we smooth the Stokes $Q_0$
and $U_0$ maps with a Gaussian beam of width $\sigma$. This results in
maps of $Q$ and $U$ which are $\sigma$ times oversampled and
which now can be used like observational maps. A noise level of
$0.05\,Q,U_{rms}$ is added to the Stokes parameters, which
corresponds to the noise in the high signal-to-noise regions in the
observations. $P$ and $\phi$ are calculated according to 
Eqs.~(\ref{e:polint}) and~(\ref{e:polang}). From the angle maps
corresponding to the five observational wavelengths we finally can
determine RM$_{fit}$ as
\begin{equation}
  \phi = \RM_{fit}\lambda^2.
\end{equation}
We apply the same criteria to select reliable fits as in the
observations (see end of \S\ref{s:obs}).

\begin{figure*}
  \centering
  \includegraphics[width=0.8\textwidth]{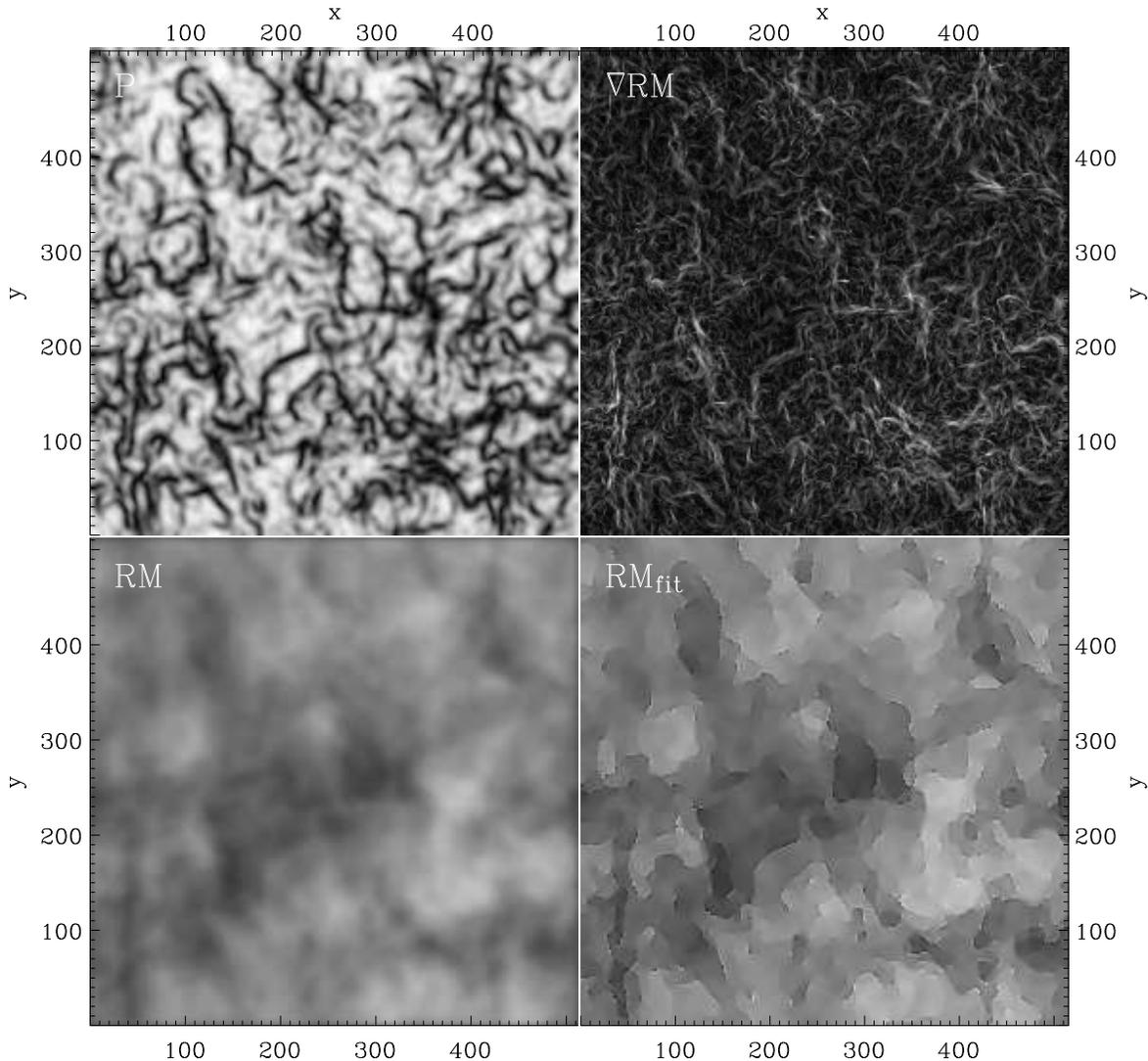}
  \caption{\label{f:pirmgrm}Polarized intensity $P$ ($4\times 10^{-3}
          < P < 0.97$), gradient of RM, $|\nabla\RM|$ ($1\times
          10^{-3}< |\nabla RM| < 4.9$~rad~m$^{-2}$~pixel$^{-1}$)
          and numerical and fitted RM (scaled identically between
          $-6.8 < RM < 22.32$~rad~m$^{-2}$) for model $B4$ at
          smoothing $\sigma=4$, four times oversampled. Note that
          depolarized (black) canals appear at the locations of large
          (white) gradients. These regions reoccur in anomalous
          $\RM_{fit}$, leading to fringes in the fitted RM map.}
\end{figure*}

Fig.~\ref{f:pirmgrm} gives an example for model $B4$ at smoothing
$\sigma=4$.
%****************************************************
%*****************************************************
$P$ shows elongated regions of depolarization, not directly related to
features in the RM map. However, the regions of low polarized
intensity $P$ occur at locations of large gradients in RM. These
structures reappear in the fitted RM, since regions of low $P$ are
noise-dominated and thus do not allow an accurate fit. The other
models show similar characteristics: elongated depolarized
regions which are one beam wide are present in all models, and
the RM derived from smoothed data is patchy and shows sub-beam scale
anomalies. However, a larger width of the RM distribution $\Delta\RM$
induces more depolarization, so that larger depolarized regions 
exist in those models. For further distinction between models, see
Sect.~\ref{s:error}.

%====================================================
\section{Depolarization canals\label{s:beamdepol}}

All high-resolution observations of the polarized intensity of the
diffuse synchrotron background exhibit elongated, one-beam-wide
``canals'' of very low or zero $P$, where the radiation is
depolarized (e.g Duncan et al.\ \cite{DHJ1997}, \cite{DRR1999}, Gray
et al.\ \cite{GLD1999}, Uyan\i ker et al.\ \cite{UFR1999}, 
Gaensler et al.\ \cite{GDM2001}, Haverkorn et al.\ \cite{HKB03a},
\cite{HKB03b}). 
The polarization angle always changes by 90$\degr$ across
such a canal (Haverkorn et al.\ \cite{HKB2000}). Two possible
explanations exist for the depolarization canals (see e.g.\ Burn
\cite{BUR1966}, Sokoloff et al.\ \cite{SBS1998}):  
\begin{enumerate}
\item beam depolarization: if the gradient in RM $\nabla\RM = (2n+1)\,
  \pi/2$ within one beam, the radiation within the beam is
  depolarized. This mechanism would explain the fact that canals do
  not shift in position with frequency. However, it requires a medium
  with steep gradients in RM and/or intrinsic angle $\phi_0$ within
  one beam, elongated perpendicularly to the gradient.
\item depth depolarization: a uniform synchrotron emitting
  magneto-ionic medium can cause complete depolarization for
  $|\RM| = n \pi$. However, in this explanation the
  depolarization canals shift in position with frequency, and a
  uniform medium (across a major part of the path length) is needed
  for complete depolarization. 
\end{enumerate}
Haverkorn et al. (\cite{HKB04b}) argue that in their observations (one
of which is used here), beam depolarization due to RM gradients is the
dominant mechanism causing depolarization canals. This does not in any
way mean that depth depolarization is not important in the rest of the
medium. For observations made at 350~MHz, only a modest gradient
in RM across a beam ($\sim 2.1$~rad~m$^{-2}$) can create a canal,
whereas much larger gradients ($\sim 34$~rad~m$^{-2}$) are needed to
form a canal in 1.4~GHz observations. For an interpretation of canals
at 1.4~GHz in terms of depth depolarization, see Shukurov \&
Berkhuijsen (\cite{SAB2003}).
 
In the numerical models discussed in Sect.~\ref{s:nummod},
depolarization canals are visible as well (see top left map in
Fig.~\ref{f:pirmgrm}), which are caused by beam depolarization.
 
\begin{figure}
  \centering
  \includegraphics[width=0.5\textwidth]{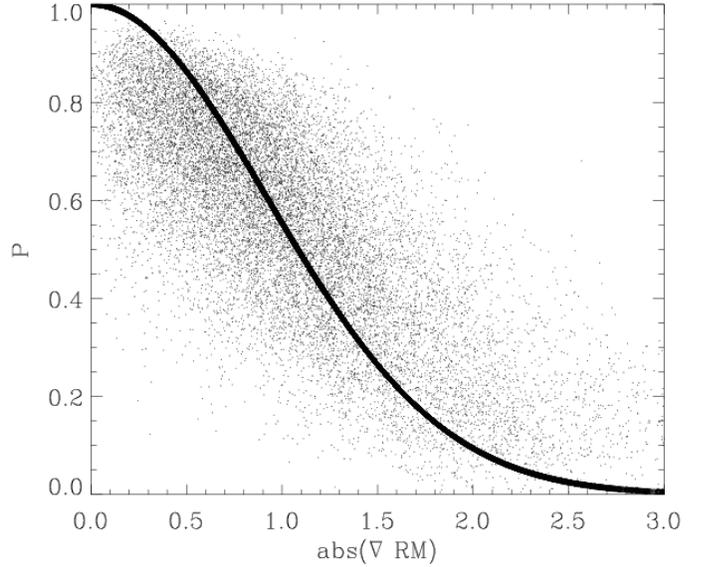}
  \caption{Absolute value of the gradient in original RM
  in the simulations $|\nabla RM|$ against polarized intensity $P$ of
  the smoothed maps with smoothing $\sigma = 4$ for each pixel. The
  curve corresponds to Eq.~\ref{e:canalpolarization}.}
  \label{f:scatter}
\end{figure}

At first sight, there is no obvious connection between the locations
of the canals and RM (left hand maps in Fig.~\ref{f:pirmgrm}). However,
the magnitude of the gradient in RM (top right map in
Fig.~\ref{f:pirmgrm}) is obviously correlated to the position of the
canals. This is demonstrated more quantitatively in
Fig.~\ref{f:scatter}, which shows a scatter plot of $P$ against the
magnitude of the gradient in RM with smoothing $\sigma=4$ for each
independent ``beam''. The solid line denotes the theoretically expected
depolarization in a beam due to a resolved gradient $d\RM/dr$ in RM
(Sokoloff et al.\ \cite{SBS1998})
\begin{equation}
  \frac{P}{P_0} = \mbox{exp} \left[ -\frac{1}{\ln 2} \left(
  \frac{d\RM}{dr} \right)^2 \lambda^4 \right]
  \label{e:canalpolarization}
\end{equation}
where $P_0$ is the polarized intensity for constant RM within the
beam. The polarized intensity decreases almost to zero where the
gradient in RM is such that $\Delta\phi \approx \pi/2$.

\begin{figure*}
  \centering
  \includegraphics{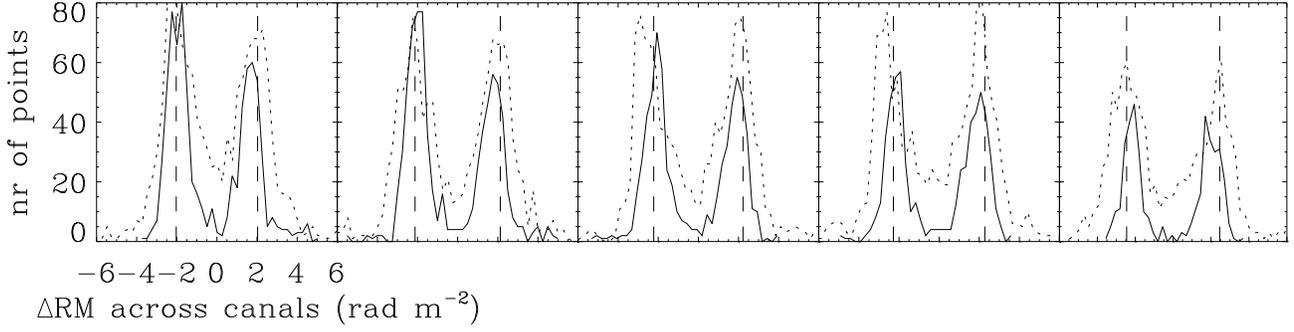}
  \caption{Distribution of difference in RM across a canal, $\Delta
  \RM$. From left to right, canals are selected in wavelengths of 
  88cm, 86cm, 84cm, 83cm and 80cm, respectively. The solid line denotes
  canals in $P$ maps smoothed by $\sigma = 4$, the
  dotted lines are for a smoothing $\sigma = 8$. The dashed lines give
  the positions where $\Delta\RM \lambda^2 = -\pi/2, \pi/2$.}
  \label{f:drmcanals}
\end{figure*}

We selected ``canal-like beams'' by comparing $P$ in a certain point
to the average $P$ of two diametrically opposed neighboring beams. If
$P$ in the central beam was less than 20\% of the averaged $P$ of the
neighboring beams, the beam was considered a canal. Note that this
selection criterion does not use any information about the length
of the canals. Fig.~\ref{f:drmcanals} shows the distribution of the
gradient in RM only at canals, defined according to the above
process. The five plots show canals defined in five wavelengths, and
the solid and dotted lines represent smoothings $\sigma = 4$ and~$8$,
respectively. The dashed lines denote the value of $\nabla$RM 
at the location where $\Delta\phi = -\pi/2, \pi/2$. 
Clearly, the gradient in RM is peaked at the value where $\Delta\phi =
\pi/2$. The decline in number of canal pixels with frequency can
be explained by the fact that the width of the polarization angle
distribution decreases with increasing frequency for a given RM
distribution.

\begin{figure}
  \centering
  \includegraphics[width=0.45\textwidth]{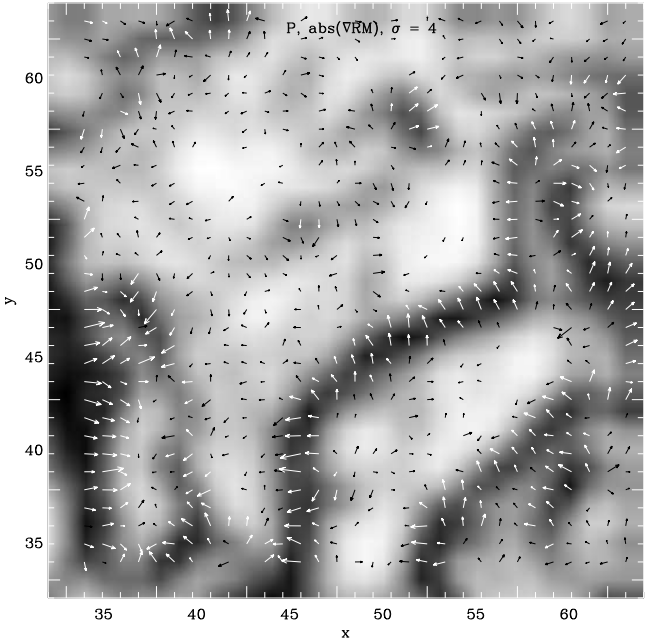}
  \caption{Part of the polarized intensity map at smoothing $\sigma = 4$
  in grey scale, with strength and direction of the gradient in
  RM superimposed as vectors.}
  \label{f:gradmap}
\end{figure}

So the gradient of RM in the canals has the right magnitude to cause
beam depolarization (Fig.~\ref{f:drmcanals}). Since the canals exist
over many beams, the most likely direction of this gradient is
perpendicular to the direction of the canals. In Fig.~\ref{f:gradmap}
we show $P$ in grey scale, with vectors of the RM gradient
superimposed, demonstrating that indeed the gradients in RM are
directed perpendicular to the canals.

Detailed analysis of shape, direction and origin
of RM gradients and the implications on the medium will be discussed
in a forthcoming paper. However, we give a simple estimate here of the
required changes in magnetic field strength $B$ and/or electron
density $n_e$ to produce large enough gradients. For $\lambda=0.80$~m,
$\Delta\RM = 2.45$~rad~m$^{-2}$ in order to  get an angle change of
$\Delta\phi=\pi/2$ within a beam. If the structures in
$\nabla$RM are filaments at an average distance of 300~pc, their width
would be 0.4~pc. The necessary $\Delta$RM requires an enhancement of
$n_e$ and $B$ of a factor 3 to 4, for ISM parameters as in e.g. model
$B2$ (see Table~\ref{tab:nummodels}). 
On the other hand, if the structures are sheetlike, and have an extent
along the line of sight of a few parsecs, then an increase in electron
density and magnetic field of 50\% is sufficient, or an increase in
either $n_e$ or $B$ of 100\%. However, the unpublished H$\alpha$
measurements discussed above (Madsen, private communication) give an
upper limit to small-scale structure in emission measure EM~=~$\int
n_e^2\; dl \approx 0.3$~cm$^{-6}$~pc, which corresponds to a maximum
$\Delta n_e \approx 0.02$~cm$^{-3}$. Therefore, the gradients in RM
are probably mainly caused by magnetic field changes, and could
originate from shocks, local narrow magnetic field reversals or other
sheet-like structures in density and/or magnetic field.

Note that the canals in the numerical models are much shorter and 
less ordered than the ones in the observations. This is because in the
numerical models we can introduce dynamical structure
self-consistently only up to the domain size. Ideally, we would like
to use subframes from the numerical models to be able to include
larger scale structure in the models, so that modeling of canals
across a significant part of the field would be possible. However,
the available resolution is too low for this to be feasible. 

Elongated narrow gradients in magnetic field and/or density have also
been seen in other numerical simulations (e.g. Schekochihin et
al.\ \cite{SMC2002}, Cho et al.\ \cite{CLV2002}). 

Concluding, in the numerical models discussed here narrow and
elongated gradients in RM exist of large enough magnitude to cause
beam depolarization in the modeled medium. Taking into account other
observational evidence (Haverkorn et al.\ \cite{HKB04b}), this leads
us to conclude that the narrow and deep depolarization canals as seen
in the observations are most likely caused by beam
depolarization due to RM gradients. 

%***************************************
%****************************************
%***************************************
%****************************************
%***************************************
%****************************************

%====================================================
\section{Reliability of RM determinations\label{s:error}}

\begin{figure}
  \centering
  \includegraphics[width=0.5\textwidth]{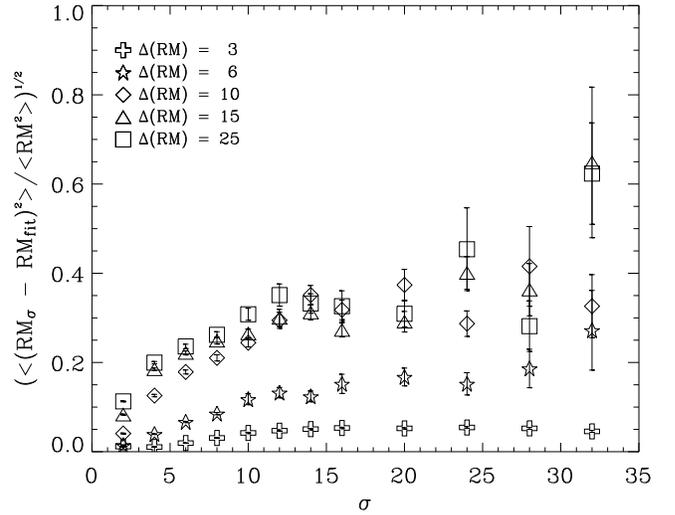}
  \caption{Deviation of RM, $\delta$RM, calculated from smoothed $Q$
  and $U$ maps from original RM, relative to original RM, against
  smoothing $\sigma$ (see Eq.~\ref{e:rmerror}). Different symbols
  denote results for models $B1$ to $B5$ with a given width of
  the original RM distribution. Only independent beam positions with
  $(P > 5S/N) \,\,\wedge\,\, (\chi^2 < 2)$ are used to determine
  $\delta$RM.}
  \label{f:errorrmfitsel}
\end{figure}

The beam depolarization raises the issue of how reliable the RM 
determinations are. Fig.~\ref{f:errorrmfitsel}
quantifies the effect of limited resolution on the 
values of the computed RM. We determine the normalized mean
quadratic difference between fitted and original smoothed RM as 
\begin{equation}
  \delta\RM = \left(\frac{\langle(\RM_\sigma-\RM_{fit})^2\rangle}
                     {\langle\RM^2\rangle}\right)^{1/2}.
  \label{e:rmerror}
\end{equation}
where RM is the original RM and RM$_\sigma$ is RM smoothed over
Gaussian beams of width $\sigma$. We have chosen for a
$\delta\RM$ defined by Eq.~(\ref{e:rmerror}) for the following
reason. The goal of the equation is to compare the fitted RM to the
``real'' RM in the modeled medium. On scales below the beam size, the
fitted RM cannot show any structure that represents structure in the
medium by definition. Therefore, if $RM_{fit}$ would be compared to
the RM as directly derived in the model, $RM_{fit} - RM$ would give
large values which are not the result of the smoothing, but only due
to the fact that the beam size is incorporated in the determination of
$\RM_{fit}$, but not in that of RM. Therefore, we chose to compare
$\RM_{fit}$ to an RM which is smoothed over a Gaussian beam, assuming
that the smoothed RM approximates the physical RM in the model if
viewed only on scales larger than the beam size. In addition, the
subsequent normalizing to the distribution width of the original {\em
unsmoothed} RM rules out any spurious effects on the difference
estimates due to averaging. In this way, $\delta\RM$ is a measure of 
how accurately the RM computed from $Q$ and $U$ data with a finite
beam width approximates the physical RM in the medium.

Fig.~\ref{f:errorrmfitsel} shows the deviations $\delta$RM for all 
independent beams satisfying the criteria for good fits 
(see end of \S\ref{s:obs}). For a width of the RM distribution
$\Delta RM \approx 20$ rad~m$^{-2}$ and a smoothing of $\sigma = 4$
(values which are typical for the observations by Haverkorn et
al. \cite{HKB03a}, \cite{HKB03b}) we would expect from
Fig.~\ref{f:errorrmfitsel} an error of $\sim 20\%$ in the fitted
RM. However, as we will see below, this error estimate is an upper
limit.
%***************************************
%***************************************

\begin{figure}
  \centering
  \includegraphics[width=0.45\textwidth]{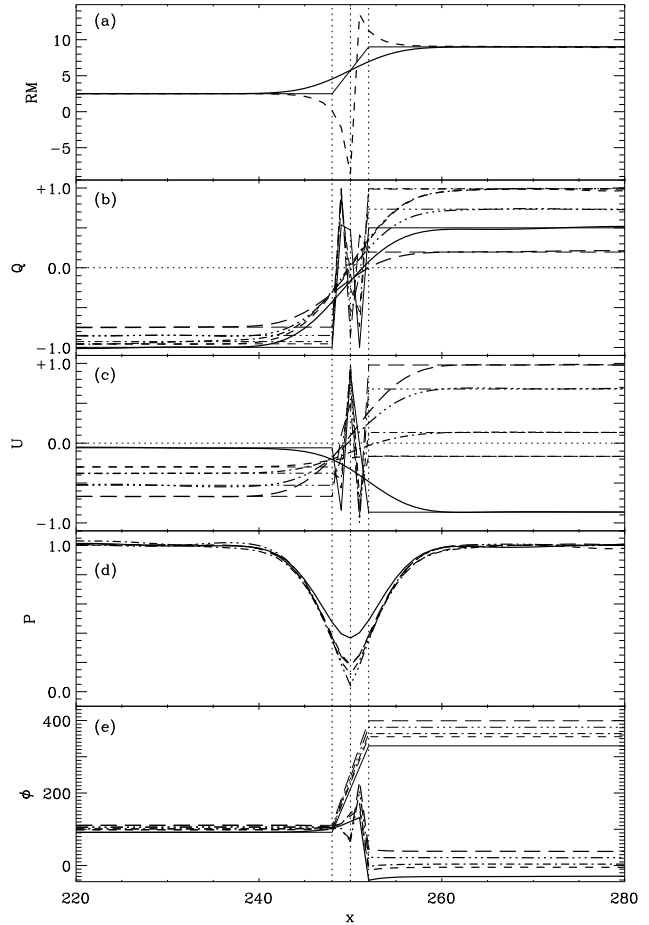}
  \caption{Details of an idealized canal caused by sign changes
           in Stokes $Q$ (see text). (a): RM against spatial
           coordinate $x$. The original RM (thin line) has a gradient
           between the two vertical dotted lines, leading to a
           smoothed RM$_s$ (thick solid line) and a fitted RM$_{fit}$
           (thick dashed line) showing an overshoot, as in the
           observations. Smoothing (``beam'') width corresponds to
           $\Delta x = 4$. $\Delta$RM as used in the text is the
           change of RM over the gradient, in this case $\Delta$RM
           $=6.5$. (b)--(d): Stokes $Q$, $U$ and polarized intensity
           $P$ for all wavelengths before smoothing (thin lines) and
           after smoothing (thick lines): Strong oscillations of the 
           unsmoothed $Q$ and $U$ within the gradient region of RM
           (see Eqs.~[\ref{e:origstokesu}] \& [\ref{e:origstokesq}])
           lead to cancellation of smoothed $Q$ and $U$ within that
           region. Correspondingly, $P$ drops to nearly $0$. 
           (e): Polarization angles for all wavelengths (as denoted 
           in legend of $Q$). Thin lines show the angles derived from
           the original, unsmoothed RM, thick lines denote the angles
           derived from the smoothed $Q$ and $U$. Note that within the
           gradient region, there is no ordering of $\phi$ with
           $\lambda$.}
  \label{f:depolcanal}
\end{figure}

For  $\Delta\RM$ small enough not to lead to sign changes of $Q$ and
$U$ within a beam, the differences $\delta RM$ saturate at a
$5$\%-level with increasing beam width, since variations of RM within
a beam are limited (Fig.~\ref{f:errorrmfitsel} for $\Delta\RM=3$). 
This behavior changes profoundly if Stokes $Q$ and $U$ reverse sign
within  one beam. This is equivalent to a change in polarization angle
of 90$\degr$ which, as discussed above, leads to depolarization
canals. The effect on $\delta$RM is twofold: 
\begin{enumerate}
  \item Low $P$ means small $S/N$, thus not well defined angles, and
        thus large errors in the fits. Therefore, those fits are
        excluded by the selection criterion.
  \item Averaging introduces slight non-linearities in the angles 
        around depolarized regions, i.e.\ regions where Stokes $Q$ and
        $U$ reverse sign within one beam. These non-linearities 
        lead to an overshoot of the RM (see Fig.~\ref{f:depolcanal})
        close to the depolarized region.
\end{enumerate}

With increasing $\Delta\RM$, Stokes $Q$ and $U$ are more likely to
change sign within a beam. Beam averaging then leads to much severer
depolarization (see Fig.~\ref{f:depolcanal}). 
%***************************************
%***************************************
Depolarized regions are accompanied by systematic overshoots in
RM$_{fit}$ (Fig.~\ref{f:depolcanal}, top panel), which stem
from the non-linearities in Eq.~(\ref{e:polang}) arising when Stokes
$Q$ changes sign. (Note that in the bottom panel of
Fig.~\ref{f:depolcanal}, the polarization angles derived from
beam-averaged Stokes $Q$ and $U$ are not ordered with respect to
$\lambda$ any more in a canal, thus leading to ill-defined
fits.) The overshoots of extreme positive and negative RM are also
visible in the observational data (see Fig.~\ref{f:obs_rm}),
indicating that the depolarization canals in the data are also caused
by beam depolarization as opposed to depth depolarization.

Not all RM gradients introduce these spurious results. Far away from
the gradient, the fits retrieve the original RM. However, close to the
gradient, multiple oscillations of $Q_0$ and $U_0$ within the gradient
lead to $Q$ and $U$ close to zero over an extended region.

\begin{figure}
  \centering
  \includegraphics[width=0.5\textwidth]{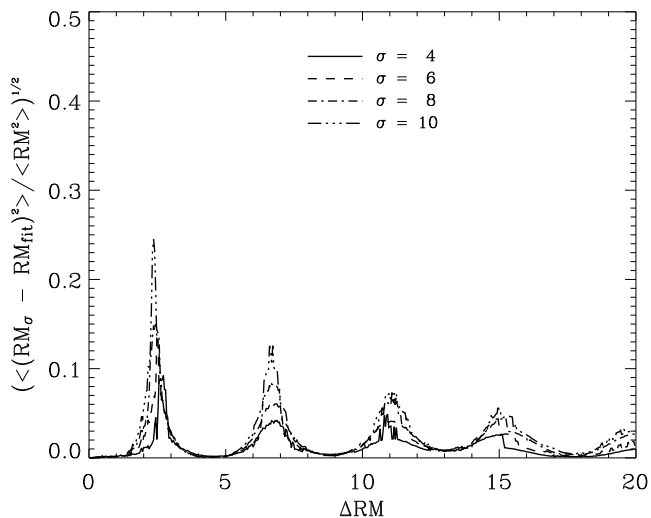}
  \caption{$\delta$RM (see Eq.~\ref{e:rmerror}) for an isolated
           RM-gradient as shown in Fig.~\ref{f:depolcanal}(a) against
           $\Delta$RM, the change of RM across the gradient. 
           The gradient width is $4$ pixels while the beam width varies
           between $4$ and $10$ pixels. Fitted RMs are derived in the
           same way as for the models, selected for $(P > 5S/N)
           \,\,\wedge\,\, (\chi^2 <2)$ and independent beams. Note the
           periodicity $T_{\mbox{\tiny{RM}}} \approx 4.4$ in
           $\delta$RM, corresponding to an angle change of $\pi$. The
           first spike occurs at an angle difference of $\pi/2$, the
           canal of 0th order.} 
  \label{f:drmperiodselect}
\end{figure}

For one isolated RM-gradient as shown in Fig.~\ref{f:depolcanal}a (see
figure caption), the differences $\delta$RM selected for reliable fits
behave with the change in RM across the canal $\Delta$RM as shown in
Fig.~\ref{f:drmperiodselect}. The first spike corresponds to a 0th
order canal with an angle difference of  $\pi/2$. Higher-order canals
have angle differences of $(2n+1)\, \pi/2$, as discussed above. The RM
field from the numerical models is a combination of all $\Delta RM$s
given in Fig.~\ref{f:drmperiodselect}.

%***************************************
%***************************************

Note that the fitted RM$_{fit}$ deviates from the original RM most 
for those RM gradients where canals occur. In between the spikes,
at positions where canals are not near, the difference $\delta\RM$
drops nearly to zero, i.e.\ a finite beam width has almost no effect
on the RM determination. Therefore, $\delta\RM$  is dominated by
values around canals where $P > 5$~S/N (needed to be selected as
reliable RM).

From Fig.~\ref{f:drmperiodselect} we conclude that the difference
estimates $\delta$RM for $\Delta\mbox{RM} > 3$~rad~m$^{-2}$ as given
in  Fig.~\ref{f:errorrmfitsel} are indeed too pessimistic, since they
are dominated by the regions around canals, that are influenced by
the canals but still fall within the definition of ``reliable RMs''. 
For smooth regions, or regions not affected by canals, we expect the
RMs determined from the observations to be accurate within $5$\% of
$\Delta$RM, corresponding to an error of $0.15$ rad~m$^{-2}$ for the
observations mentioned here. The overall error will increase with
increasing number of canals, which is why Fig.~\ref{f:errorrmfitsel}
generally shows larger $\delta\RM$ than Fig.~\ref{f:drmperiodselect}. 

%====================================================
\section{Conclusions}
\label{s:conc}

Multi-frequency radio observations of the Galactic synchrotron 
background allow us to probe the structure of the magneto-ionized
component of the ISM from RM maps.
However, the quality of the RM determinations depends strongly
on the degree of local depolarization due to intrinsic structure
in the ISM, and thus in the RM. We investigated the effect of beam
depolarization on the quality of the fitted RMs with the help
of numerical models. We find:
\begin{enumerate}
  \item The models show depolarization canals due to beam
        depolarization across sharp gradients in RM. This suggests
        that narrow gradients of a few rad~m$^{-2}$ can be a common
        feature in a magneto-ionized medium. Combined with other
        observational evidence, we therefore conclude that the
        depolarization canals seen in our 350~MHz observations
        are caused by beam depolarization due to RM gradients.
        While the width of the canals is an observational
        artifact, their presence and length mirror underlying
        structure in the Faraday-rotating medium.
  \item RM determinations close to canals can be incorrect by several
        rad~m$^{-2}$, depending on the absolute RM difference between
        both edges of the canal. This is a consequence of a
        non-linearity in $\phi(\lambda^2)$ introduced by finite beam
        width. 
  \item RM determinations in smooth regions or in regions separated
        from canals by at least a beam are accurate to
        $0.05\Delta\RM$, where $\Delta\RM$ is the width of the RM
        distribution, or $\delta\mbox{RM} = 0.15$ rad~m$^{-2}$ for
        the observations mentioned here.
\end{enumerate}

In future research, improved models will enable us to investigate what
structures in the ISM actually cause the elongated gradients of RM,
and will serve as a calibration what structure information we can
extract from observational maps.

\begin{acknowledgements}
The authors wish to thank Ellen Zweibel, Peter Katgert, 
Mordecai-Mark Mac Low and Enrique V\'azquez-Semadeni for enlightening
and  invigorating discussions. We are grateful to P.~S. Li and
M.~L. Norman for making the  $512^3$ available, and thank Wolfgang
Reich for useful and critical comments.
Computations were performed on the SGI Origin 2000 of the National
Center for Supercomputing Applications (NCSA) at Urbana-Champain.
The Westerbork Synthesis Radio Telescope is operated by the
Netherlands Foundation for Research in Astronomy (ASTRON) with
financial support from the Netherlands Organization for Scientific
Research (NWO). The project originated during the Astrophysical
Turbulence Program at the KITP at  University of Californa, Santa
Barbara in 2000 (NSF grant PHY94-07194). 
M.H. is supported by NWO grant 614-21-006. F.H. is supported by a
Feodor-Lynen fellowship of the Alexander-von-Humboldt Foundation and
by NSF grant AST-0098701. 

\end{acknowledgements}


\begin{thebibliography}{}
\bibitem[1995]{ARS1995} 
        Armstrong, J.~W., Rickett, B.~J., \& Spangler, S.~R. 1995, 
        \apj, 443, 209 
\bibitem[2002]{DAM2002}
        de Avillez, M.~A., \& Mac Low, M.-M. 2002, \apj, 581, 1047 
\bibitem[1966]{BUR1966} 
        Burn, B. J. 1966, \mnras, 133, 67
\bibitem[2002]{CLV2002} 
        Cho, J., Lazarian, A., \& Vishniac, E.~T. 2002, \apj, 564, 291
\bibitem[2003]{CLV2003} 
        Cho, J., Lazarian, A., \& Vishniac, E.~T. 2003, \apj, 595, 812
\bibitem[1999]{CRU1999} 
        Crutcher, R.~M. 1999, \apj, 520, 706 
\bibitem[1997]{DHJ1997}
        Duncan, A. R., Haynes, R. F., Jones, K. L., \& Stewart,
        R. T. 1997, \mnras, 291, 279
\bibitem[1999]{DRR1999}
        Duncan, A. R., Reich, P., Reich, W., \& F\"urst, E. 1999, \aap,
        350, 447 
\bibitem[2001]{GDM2001} Gaensler, B. M., Dickey, J. M.,
        McClure-Griffiths, N. M., Green, A. J., Wieringa, M. H., \&
        Haynes, R. F. 2001, \apj, 549, 959 
\bibitem[1999]{GLD1999} Gray, A.~D., Landecker, T.~L., Dewdney, P.~E.,
        Taylor, A.~R., Willis, A.~G., Normandeau, M. 1999, \apj, 514,
        221 
\bibitem[2000]{HKB2000} 
        Haverkorn, M., Katgert, P., \& de Bruyn, A. G. 2000, \aap,
        356, L1
\bibitem[2003a]{HKB03a} 
        Haverkorn, M., Katgert, P., \& de Bruyn, A. G. 2003a, \aap,
        403, 1031  %auriga 
\bibitem[2003b]{HKB03b} 
        Haverkorn, M., Katgert, P., \& de Bruyn, A. G. 2003b, \aap,
        404, 233 % ring
\bibitem[2004a]{HKB04a} 
        Haverkorn, M., Katgert, P., \& de Bruyn, A. G. 2004a,
        submitted to \aap  %ch3b, model 
\bibitem[2004b]{HKB04b} 
        Haverkorn, M., Katgert, P., \& de Bruyn, A. G. 2004b,
        submitted to \aap  %ch3a, depol 
\bibitem[2001]{HZM2001} 
        Heitsch, F., Zweibel, E.~G., Mac Low, M.-M., Li, P.~S.,
        Norman, M.~L. 2001, \apj, 561, 800 
\bibitem[1981]{LAR1981}
        Larson, R.~B. 1981, \mnras, 194, 809 
\bibitem[2000]{LNM2000} 
        Li, P.~S., Norman, M.~L., Heitsch, F., \& Mac Low, M.-M. 2000, 
        \baas, 197, 502 
\bibitem[2004]{LNM2004} 
        Li, P.~S., Norman, M.~L., Mac Low, M.-M., \& Heitsch, F. 2004, 
        \apj, accepted
\bibitem[1999]{MAC1999}
        Mac Low, M.-M.  1999, \apj, 524, 169
\bibitem[2003]{MAK2003}
        Mac Low, M.-M., \& Klessen, R. 2003, ARAA, in press, 
        astro-ph/0301093
\bibitem[1998]{MAB1998}
        Minter, A., \& Balser, D. S. 1998, LNP, 506, 543
\bibitem[2000]{NOR2000}
        Norman, M.~L. 2000, Revista Mexicana de Astronomia y
        Astrofisica Conference Series, 9, 66  
\bibitem[1991]{REY1991}
        Reynolds, R.~J. 1991, \apj, 372, 17L
\bibitem[2002]{SMC2002}
        Schekochihin, A.~A., Maron, J.~L., Cowley, S.~C., \&
        McWilliams, J.~C. 2002, \apj, 576, 806
\bibitem[1999]{SEB1999} 
        Sellwood, J.~A.,~\& Balbus, S.~A. 1999, \apj, 511, 660 
\bibitem[2003]{SAB2003} 
        Shukurov, A., \& Berkhuijsen, E. M. 2003, \mnras, 342, 496
\bibitem[1998]{SBS1998} 
        Sokoloff, D. D., Bykov, A. A., Shukurov, A., Berkhuijsen,
        E. M., Beck, R., \& Poezd, A. D. 1998, \mnras, 299, 189 
\bibitem[1999]{UFR1999} Uyan\i ker, B., F\"urst, E., Reich, W., Reich,
        P., \& Wielebinski, R. 1999, \aaps, 138, 31
\end{thebibliography}
\end{document}